\documentclass[3p,times]{elsarticle}

\usepackage{ecrc}

\volume{00}

\firstpage{1}

\journalname{Physics Letters A}

\runauth{}

\jid{pla}

\jnltitlelogo{Physics Letters A}

\CopyrightLine{2011}{Published by Elsevier Ltd.}

\usepackage{amssymb}

\biboptions{compress}

\usepackage[figuresright]{rotating}

\begin{document}

\begin{frontmatter}



\dochead{}

\title{String networks with junctions in competition models}


\author[cfa-up,dfa-up]{P. P. Avelino}
\address[cfa-up]{Centro de Astrof\'{\i}sica da Universidade do Porto,
Rua das Estrelas, 4150-762 Porto, Portugal}
\address[dfa-up]{Departamento de F\'{\i}sica e Astronomia, Faculdade
de Ci\^encias, Universidade do Porto, Rua do Campo Alegre 687,
4169-007 Porto, Portugal}

\author[df-ufpb]{D. Bazeia}
\address[df-ufpb]{Departamento de F\'{\i}sica, Universidade Federal da
Para\'{\i}ba, 58051-970 Jo\~ao Pessoa, PB, Brazil}

\author[df-ufpb]{L. Losano}

\author[ua,ect-ufrn]{J. Menezes}  
\address[ect-ufrn]{Escola de Ci\^encias e Tecnologia, Universidade
Federal do Rio Grande do Norte\\ Caixa Postal 1524, 59072-970 Natal,
RN, Brazil}
\address[ua]{Institute for Biodiversity and Ecosystem
Dynamics, University of Amsterdam, Science Park 904, 1098 XH
Amsterdam, The Netherlands} 

\author[dfi-uem]{B. F. de Oliveira}  
\address[dfi-uem]{Departamento de F\'{\i}sica, Universidade Estadual
de Maring\'a, Av. Colombo, 5790, 87020-900 Maring\'a, PR, Brazil}

\begin{abstract}
In this work we give specific examples of competition models, with six
and eight species, whose three-dimensional dynamics naturally leads to
the formation of string networks with junctions, associated with
regions that have a high concentration of enemy species. We study the
two- and three-dimensional evolution of such networks, both using
stochastic network and mean field theory simulations. If the
predation, reproduction and mobility probabilities do not vary in
space and time, we find that the networks attain scaling regimes with
a characteristic length roughly proportional to $t^{1/2}$, where $t$
is the physical time, thus showing that the presence of junctions, on
its own, does not have a significant impact on their scaling
properties.
\end{abstract}


\end{frontmatter}


\section{Introduction}
\label{sec:int}

Competition models are widely regarded as a crucial tool to understand
the mechanisms leading to biodiversity \cite{Kerr2002, Leisner2012,
Cheng2014, Daly2015189} (see also \cite{sole2006selforganization,
nowak06evolutionaryDynamicsBOOK} for a review). Although the simplest
competition models usually consider three species and allow for an
equal number of basic microscopic actions (motion, reproduction and
predation), many interesting generalizations, including further
species and more complex interaction rules, have been considered in
the literature \cite{May-Leonard, Reichenbach2007,
PhysRevLett.99.238105, Peltomaki2008, PhysRevE.83.011917,
PhysRevE.85.051903, 1742-5468-2012-07-P07014, PhysRevE.86.031119,
PhysRevE.86.036112, PhysRevE.87.032148, PhysRevE.88.022123,
PhysRevE.89.012721, PhysRevE.89.032133, PhysRevE.89.042710,
PhysRevE.90.032704, PhysRevE.90.042920, 1751-8121-47-16-165001,
Gokhale2014, Avelino2014393, 1367-2630-17-11-113033, Weber20140172,
PhysRevE.91.033009, PhysRevE.91.052135, PhysRevE.92.022820,
Roman201610}

In \cite{PhysRevE.86.031119, PhysRevE.86.036112}, a broad family of
spatial stochastic May-Leonard models with an arbitrary number of
species has been introduced. There, many interesting features of this
family of models, including the dynamics of complex networks of
spiralling patterns and interfaces, have been investigated in detail.
Recently, in \cite{Avelino2014393}, it has been shown that specific
sub-classes of this family of models may lead to the emergence of
string networks in three spatial dimensions. These strings are
associated to predator-prey interactions which occur mainly along
lines corresponding to a high concentration of enemy species. The
strings studied in \cite{Avelino2014393} do not have junctions and
their dynamics has been shown to be curvature driven.

In this paper we consider the emergence of string networks with
junctions in the context of specific spatial stochastic competition
models belonging to the general family introduced in
\cite{PhysRevE.86.031119, PhysRevE.86.036112}. We investigate the two-
and three-dimensional dynamics of these models using both stochastic
and mean field network simulations. The outline of this paper is as
follows.  In Set.~\ref{sec:family} we introduce a sub-class of
stochastic May-Leonard models allowing for the formation of string
networks with junctions in three spatial dimensions. In
Sec.~\ref{sec:sto} we investigate the stochastic evolution of two- and
three-dimensional networks in models belonging to the above sub-class
with $N=6$ or $N=8$ species. In Sec.~\ref{sec:ft} we study the
dynamics of such systems using mean field theory simulations,
considering also the particular case of the collapse of a circular
loop. In Sec.~\ref{sec:sc} we constrain the scaling parameter
$\lambda$ governing the macroscopic dynamics of these models,
considering various choices for the mobility, predation and
reproduction parameters. Finally, we conclude in Sec.~\ref{sec:end}.

\section{Family of models}
\label{sec:family}

\begin{figure}[t]
\centering
\includegraphics[scale= 0.70]{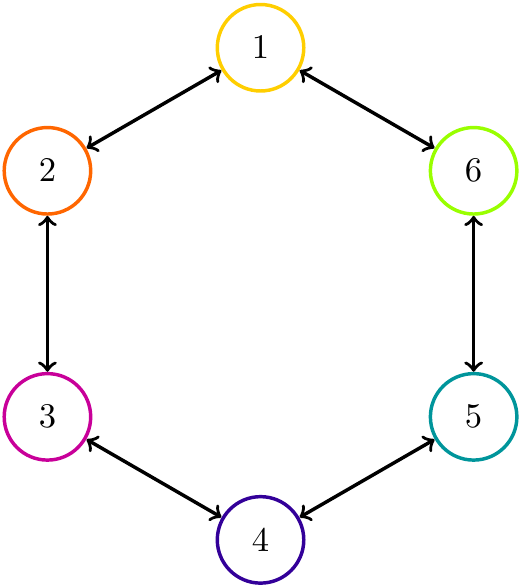}
\hspace{2cm}
\includegraphics[scale= 0.70]{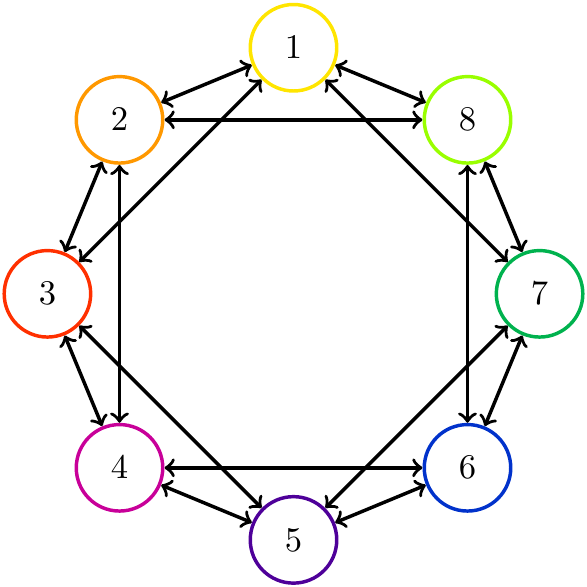}
\caption{Diagrams describing the possible predator-prey
interactions in the models with $N=6$ (left panel) and $N=8$ (right
panel) different species.}
\label{fig1}
\end{figure}

Here, we consider a sub-class of the more general family of spatial
stochastic May-Leonard models introduced in
Refs.~\cite{PhysRevE.86.031119, PhysRevE.86.036112}. We investigate
models with an even number of species $N>4$, where each species
competes with $N-4$ other species. The competition diagrams for the
simplest cases with $N=6$ and $N=8$ species are illustrated in the
left and right panels of Fig.~\ref{fig1}, respectively. The double
arrows indicate that the predation between competing species is
bi-directional. Except for the labelling of the different species, the
diagrams depicted in Fig.~1 are invariant under rotations by
$\theta_n=2\,\pi(n/N)$, with $n=0,\pm1,\pm2,\dots,\pm(N-1)$,
indicating the presence of a $Z_N$ symmetry.

In these models individuals of $N$ species are initially distributed
on square or cubic lattices with ${\mathcal N}$ sites. The different
species are labeled by $i= 1,\ldots,N$, and the cyclic identification
$i= i+ k\, N$, where $k$ is an integer, is made. The sum of the number
of individuals of the species $i$ ($I_i$) with the number of empty
sites ($I_E$) is equal to the total number of sites (${\mathcal N}$),
that is
\begin{equation}
	\sum_{i=1}^N I_i+I_E={\mathcal N}.
	\label{eq1} 
\end{equation}

At each time step a random individual (active) is chosen to interact
with one of its nearest neighbours (passive), also selected randomly.
The number of neighbours is equal to four, in the case of the
two-dimensional square lattice, and to six, in the case of the
three-dimensional cubic lattice. The unit of time $\Delta t=1$ is
defined as the time necessary for ${\mathcal N}$ interactions to occur
(one generation). The possible interactions are classified as Motion $
i\ \odot \to \odot\ i\,, $ Reproduction $ i\ \otimes \to ii\,, $ or
Predation $i\ \ (i \pm \alpha) \to i\ \otimes\,, $ where $\odot$ may
be any species ($i$) or an empty site ($\otimes$) and
$\alpha=1,\ldots,(N-4)/2$. A constant predation probability $p$
between competing species, and constant reproduction and mobility
probabilities ($r$ and $m$, respectively), common to all species, is
considered. Although this class of models is defined for a generic
even $N>4$, in this paper we shall investigate explicitly the models
illustrated in Fig.~\ref{fig1}, with $N=6$ and $N=8$ different
species.

\section{Stochastic network simulations}
\label{sec:sto}

We performed a series of stochastic network simulations of the models
with six and eight species in two and three spatial dimensions (in
square and cubic lattices, respectively). The numerical results show
that, as soon as the simulations start, individuals of the same
species, originally distributed randomly throughout the lattice, share
common spatial regions. The individuals tend not to be close to
competitors, but in the vicinity of individuals of the same species or
of other neutral species.

\begin{figure}[t]
\centering
\includegraphics[scale=0.7]{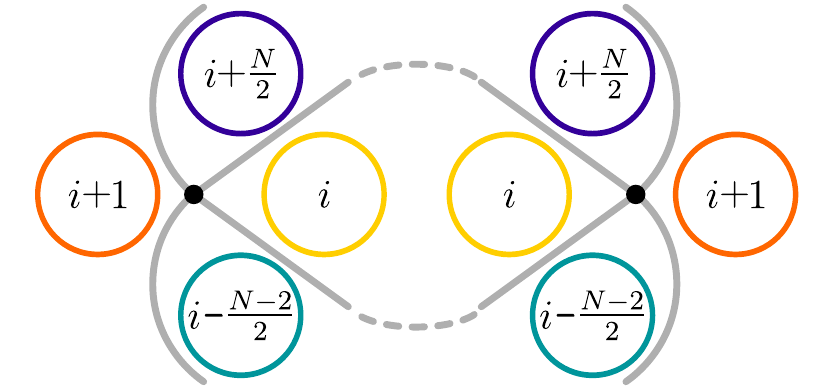}
\caption{Illustration of a defect/anti-defect pair with
four species in a model with an arbitrary number of species $N$, where
$i=1,...,\frac{N}2$.}
\label{fig2}
\end{figure}

In two spatial dimensions such spatial configurations promote the
coexistence of species which may be organized either clockwise or
counterclockwise, around roughly circular regions with a significantly
higher density of empty sites. Attacks and counter-attacks between
competing species on opposite sides of the battle cores ensure the
stability of such spatial patterns. Similar two-dimensional
arrangements of the species have been found in
Ref.~\cite{Avelino2014393}. These can be described as
defect/anti-defect configurations associated to
clockwise/counterclockwise vortex states, respectively. The average
area of the defect and anti-defect cores depends on the interaction
probabilities ($m$, $p$, $r$), but it is roughly constant in time.

In contrast with the model described in Ref.~\cite{Avelino2014393},
where every species would compete with $N-3$ distinct ones, here each
species has less competitors (each species has $N-4$ competitors and
three neutral species). As a consequence, the number of configurations
which promote coexistence among species is enlarged. In fact, while in
Ref.~\cite{Avelino2014393} all $N$ species have to be gathered around
the defect cores in order to guarantee their stability, here, a stable
defect configuration requires only four species. Defects/anti-defects
arise when the different species dispose themselves in the
clockwise/counterclockwise vortex configurations shown in Fig.
\ref{fig2} $\{i, i+\frac{N}{2}, i+1, i-\frac{N-2}{2}\}$, where
$i=1,...,\frac{N}{2}$ (note that each species $i$ does not interact
with the species $i\pm \frac{N-2}{2}$ and $i+\frac{N}{2}$). These
configurations account for $\frac{N}{2}$ different kinds of
defect/anti-defects with four species.

Consider the competing species $i$ and $i+1$ belonging to the
defect/anti-defect configuration shown in Fig. \ref{fig2}. The average
number of attacks per unit time from individuals of the outer species
$i + 1$ supersedes those from individuals of the inner species $i$.
This implies that individuals of the outer species tend to invade the
territory of the inner ones causing an approximation and annihilation
of the defect/anti-defect pair. We shall show that the
defect/anti-defect cores attract each other, having a velocity which
is, on average, inversely proportional to the distance between them.
In contrast, a pair of clockwise (or counterclockwise) defects
(anti-defects) cannot annihilate and repel each other.

Moreover, defect/anti-defect configurations with a larger number of
species may arise. In fact, defect/anti-defect configurations with
$n=4, ..., \frac{N+2}{2}$ species, in which each species competes with
$n-3$ other species, may appear in models with an even number $N > 6$
of species.

All network simulation snapshots presented throughout this paper were
obtained by assuming $m = 0.10$, $p = 0.80$ and $r = 0.10$. Although
these values were found to be adequate for visualization purposes, we
verified that many other choices of the parameters would provide
similar qualitative results.

\subsection{6 species}

\begin{figure}[t]
\centering
\includegraphics[width= 4.2cm]{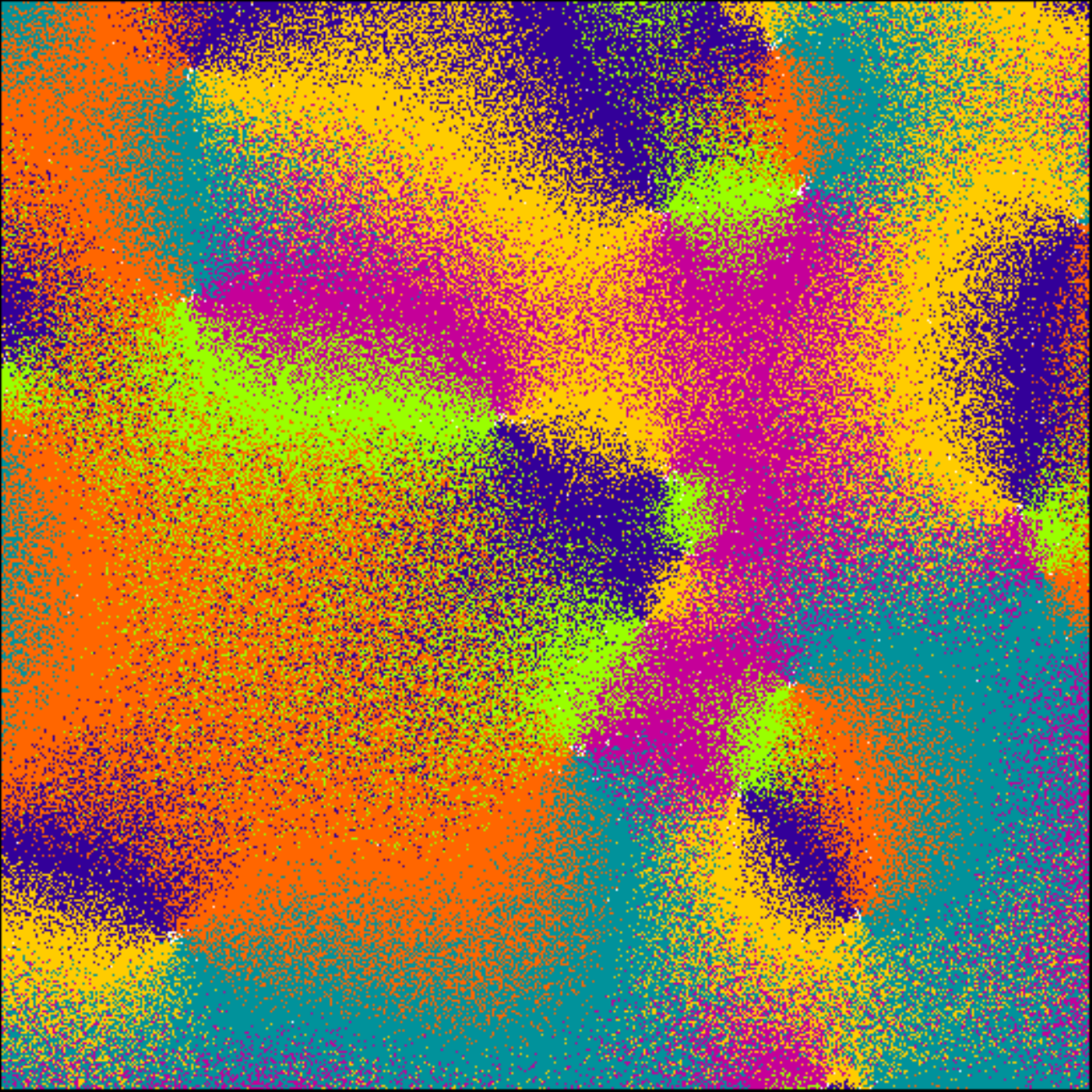}
\hspace{2cm}
\includegraphics[width= 4.2cm]{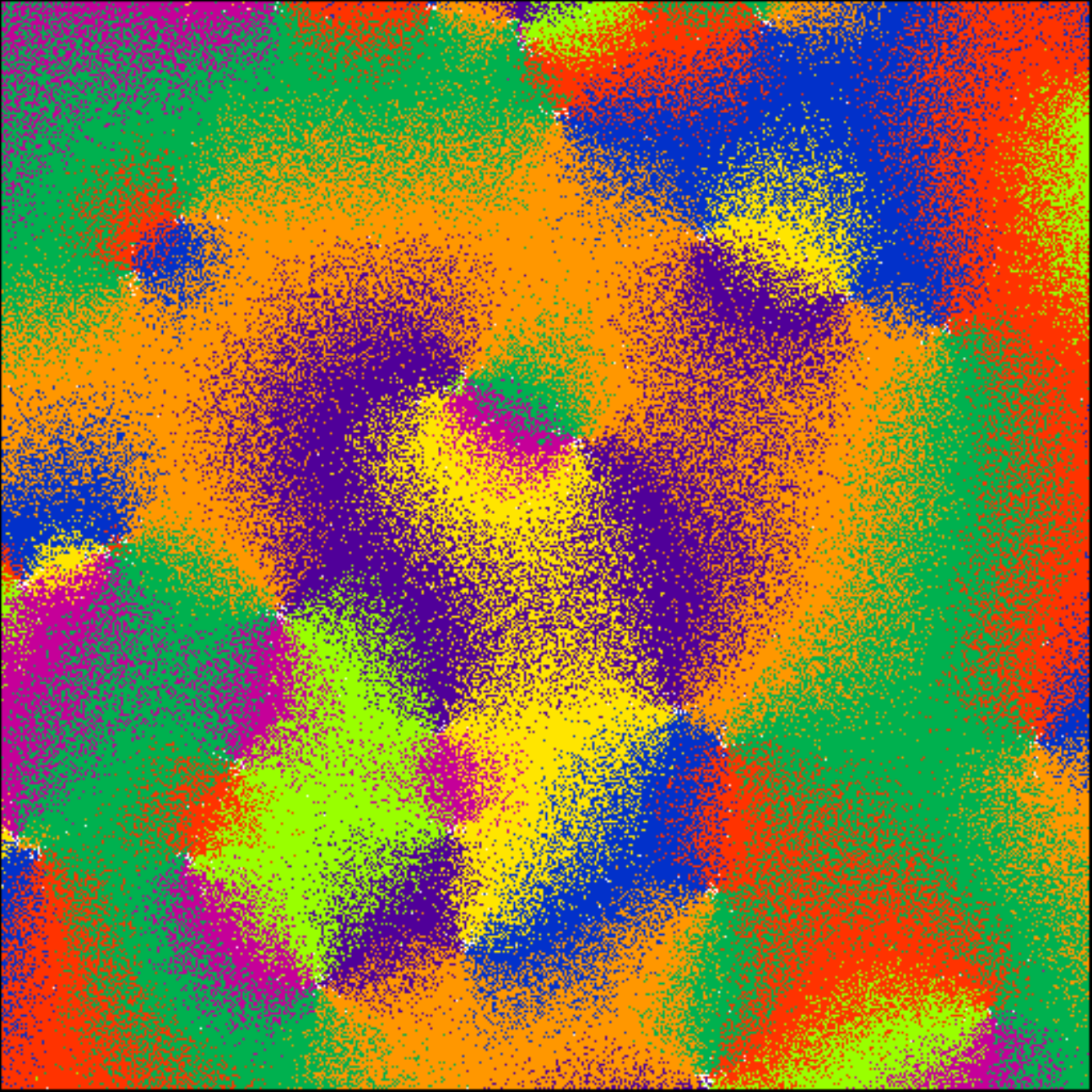}
\caption{Snapshots obtained from two-dimensional
$1024^2$ stochastic network simulations of the $N=6$ (left panel) and
$N=8$ (right panel) models. The color scheme used here is the same as
represented in Fig. \ref{fig1} for $i=1$.}
\label{fig3}
\end{figure}

Let us focus on the simplest case with $N=6$. The left panel of Fig.
\ref{fig3} shows one snapshot taken from a two-dimensional $1024^2$
stochastic network simulation with periodic boundary conditions. Each
grid point is at most occupied by one individual which belongs to the
species indicated by the same color as in Fig. \ref{fig1}. In
addition, empty spaces are represented by white dots.

The spatial patterns show that all defects involve four different
species. Most of the predator-prey interactions take place at the
defect core, which is surrounded by two pairs of competing domains,
each domain being dominated by a single species. More specifically,
one may identify three types of defects-anti/defects where domains
dominated by the species $\{ i, i-1, i+1, i-2\}$ (where $i=1,2,3$)
arrange themselves in this order (or the reverse) around the defect
(anti-defect) cores. Due to the periodic boundary conditions, the
number of defects in each simulation is equal to the number of
anti-defects. As the network evolves, the number of defects becomes
increasingly smaller due to the annihilation of defect/anti-defect
pairs. Ultimately, at large enough $t$, only one group of three
species which do not compete each other may survive (either $(1,3,5)$
or $(2,4,6)$), and the symmetry of the model is said to be
spontaneously broken.

We also run a series of $256^3$ three-dimensional stochastic network
simulations with periodic boundary conditions of the $N=6$ model. We
notice that the extension to three spatial dimensions of the dynamics
presented in the left panel of Fig. \ref{fig3} gives rises to a string
network with Y-type junctions. Such strings represent regions with a
significant larger number density of empty spaces, which appear as a
consequence of the frequent predation interaction between competing
species taking place at their core.

\begin{figure}[t]
\centering
\includegraphics[width= 4.2cm]{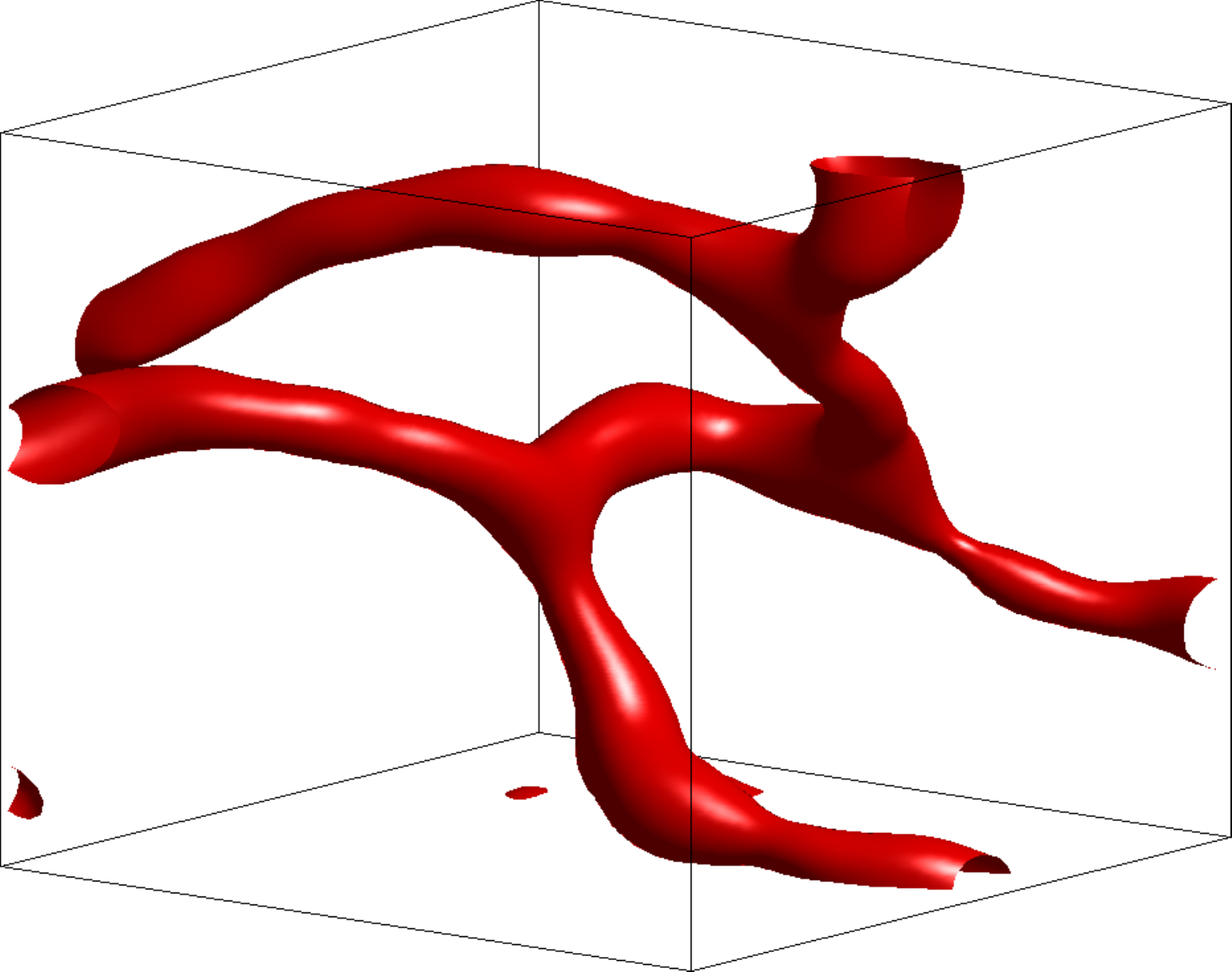}
\hspace{2cm}
\includegraphics[width= 4.2cm]{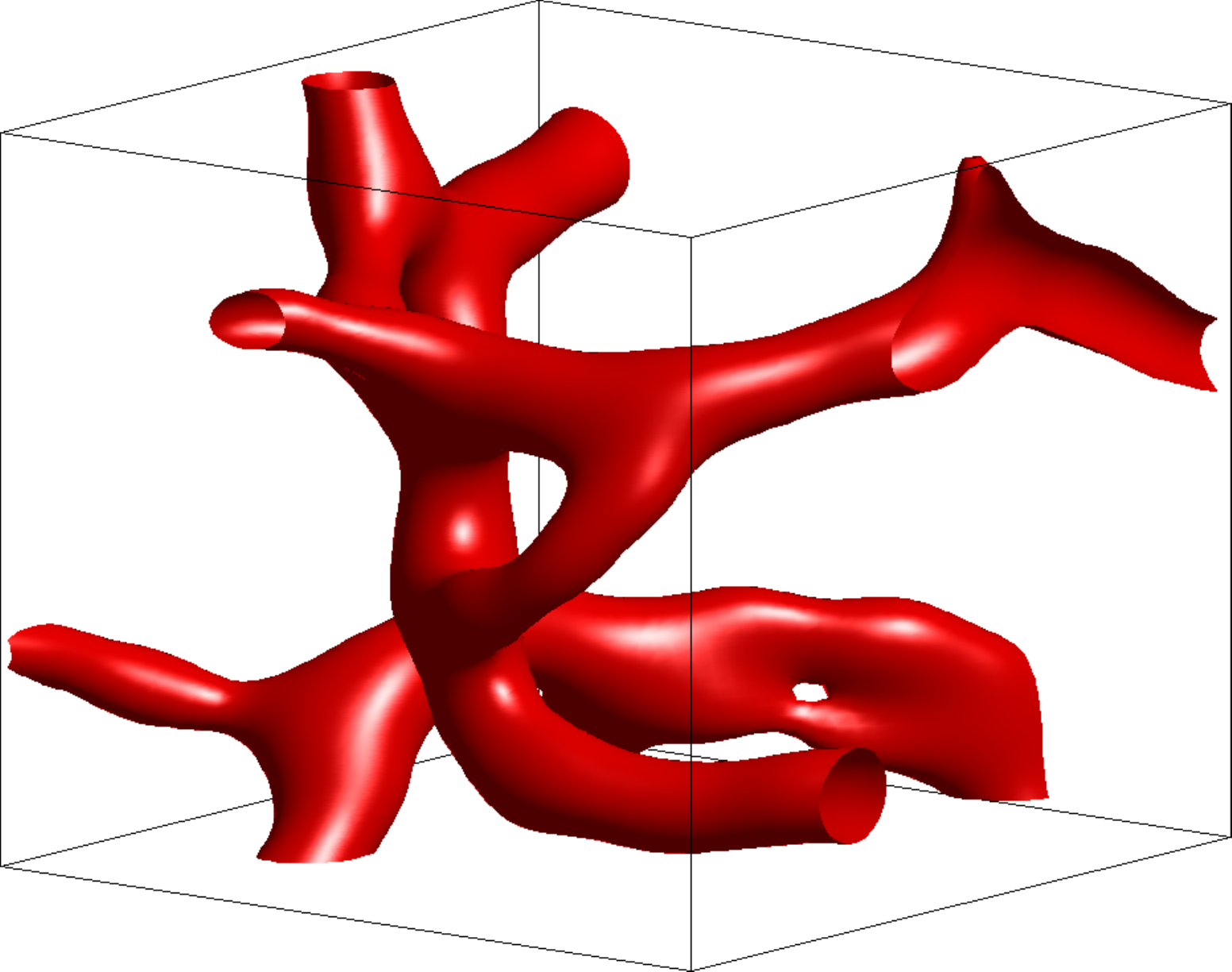}
\caption{Snapshots taken from a three-dimensional $64^3$ region of a
$256^3$ stochastic network simulation of the $N=6$ (left panel) and
$N=8$ (right panel) models.}
\label{fig4}
\end{figure}

The snapshot shown in the left panel of Fig.~{\ref{fig4}} represents a
$64^3$ region of the entire $256^3$ three-dimensional lattice. It
presents the contour plots associated to a fixed value of the density
of empty sites, which highlight the presence of a string network with
junctions (note that, in order to improve the visualization, the
number density of empty spaces has been convolved with a gaussian
filter function).

Throughout their evolution the strings intersect and intercommute
(exchange partners). This process is responsible for the production of
string loops which collapse with a characteristic velocity roughly
proportional to the loop characteristic scale. The collapse of a
string loop is curvature driven and it is associated to the existence
of a defect/anti-defect pair on any plane intersecting the loop. The
loop collapse will be further considered in Sec.~\ref{loop}.

Since any planar defect or anti-defect involves only four species,
each species joins two different strings in three spatial dimensions.
This is responsible for the appearance of Y-type junctions, defined as
the meeting point of three different strings. The stability of the
junctions is ensured by predator-prey interactions, given that each
species has competitors in two different strings.

The upper panel of Fig.~{\ref{fig5}} illustrates the different
possible configurations of a string junction in the $N=6$ model, where
$i=1,2,3$. Note that, in this model, all the species are involved in
the formation of a Y-type junctions, each one of them being associated
to two different strings.

\subsection{8 species}

The results obtained for $N=8$ are qualitatively similar to those
presented for $N=6$, but with an increased complexity associated with
the larger number of species. In this case, apart from the four
different defects/anti-defects with four species represented by
$\{i,i+4,i+1,i-3\}$ (where $i=1,..4$), there are eight
defect/anti-defect configurations composed by five species. They are
composed by the clockwise/counterclockwise disposition of the species
$\{i,i+4,i+1,i-2,i+3\}$ (where $i=1,..8$) around the defect cores. We
call the defects formed by four and five species, type I and II,
respectively.

The right panel of Fig. \ref{fig3} depicts one snapshot taken from a
$1024^2$ stochastic network simulation of the $N=8$ model, with
periodic boundary conditions. The colors indicate the species that
individuals belong to (See Fig. \ref{fig1}). White dots represent
empty sites.

In addition, the results provided by three-dimensional simulations are
presented in the right panel of Fig. \ref{fig4}. The snapshot
represents a $64^3$ region of a three-dimensional $256^3$ stochastic
network simulations with periodic boundary conditions of the $N=8$
model.
\begin{figure}[t]
\centering
\includegraphics[scale=0.70]{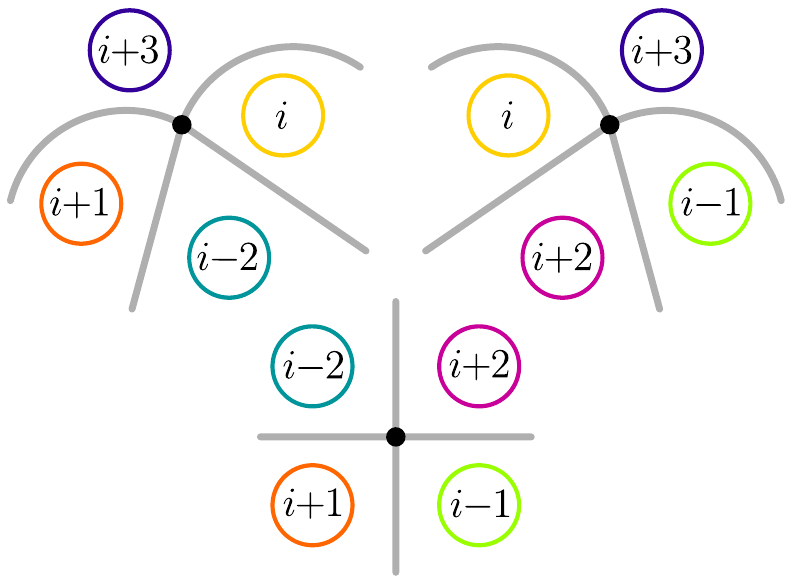}
\hspace{2cm}
\includegraphics[scale=0.70]{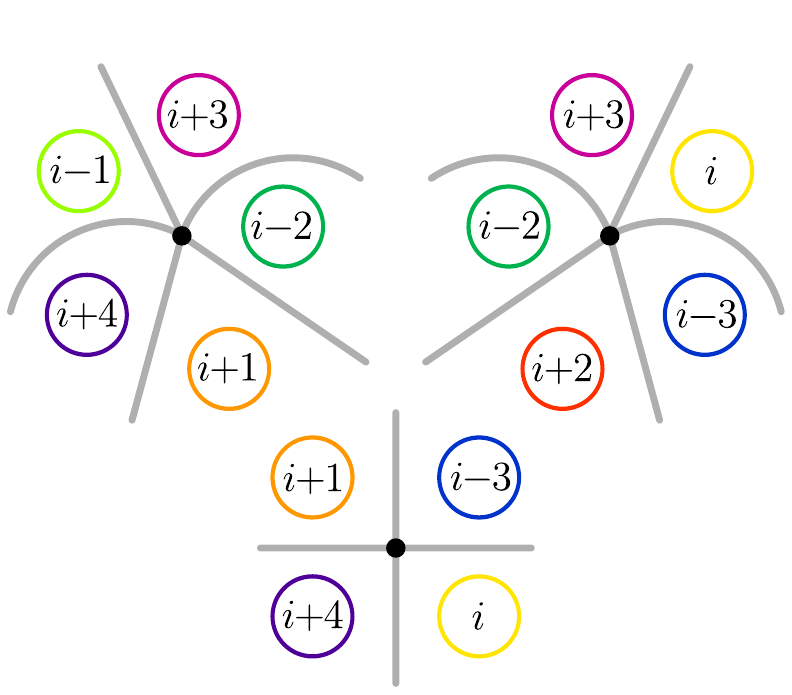}
\caption{Illustration of the formation of Y-type
junctions in the models $N=6$ (left panel) and $N=8$ (right panel).
The cross-sections of the different types of strings joining the
junctions are presented, where $i=1,...,\frac{N}{2}$.}
\label{fig5}
\end{figure}

Analogously to the $N=6$ model, the extension to three spatial
dimensions gives rises to a string network with junctions.
Nonetheless, in the $N=8$ model there are two different types of
strings, in which either four (type I) or five different species (type
II) are involved. One string of type I and two of type II are required
in order to produce the stable Y-type junctions seen in the
simulations of the $N=8$ model. The lower panel of Fig. \ref{fig5}
illustrates the formation of a Y-type junction in this model.

\section{Mean field theory simulations}
\label{sec:ft}

Let us define $N+1$ scalar fields ($\phi_0$, $\phi_1$, $\phi_2$,
$\ldots$, $\phi_N$) representing the fraction of space around a given
point occupied by empty sites ($\phi_0$) and by individuals of the
species $i$ ($\phi_i$), satisfying the constraint
\mbox{$\phi_0+\phi_1+ \ldots+\phi_{N}=1$}. For an even $N > 4$ the
mean field equations of motion
\begin{eqnarray}
{\dot \phi}_{0} &=& D \nabla^2 \phi_0 -r \phi_0 \sum_{i=1}^{N}\phi_i
+ p\, \sum_{i=1}^{N}
\,\left(\sum_{\alpha=1}^{\frac{N-4}{2}}
\phi_{i}\phi_{i+\alpha}+\sum_{\alpha=\frac{N+4}{2}}^{N-1}
\phi_{i}\phi_{i+\alpha}\right)\,,\label{1}\\ {\dot \phi}_{i} &=& D
\nabla^2 \phi_i +r \phi_0 \phi_i - p
\left(\sum_{\alpha=1}^{\frac{N-4}{2}} \phi_{i}\phi_{i+\alpha}+
\sum_{\alpha=\frac{N+4}{2}}^{N-1}
\phi_{i}\phi_{i+\alpha}\right)\,,\label{2}
\end{eqnarray}
describe the average dynamics of the models studied in the previous
section (see \cite{May-Leonard}, for more details). Here, a dot
represents a derivative with respect to the physical time and $D=2m$
is the diffusion rate.

\subsection{Two and three-dimensional numerical results}

\begin{figure}[t]
\centering
\includegraphics[width= 4.2cm]{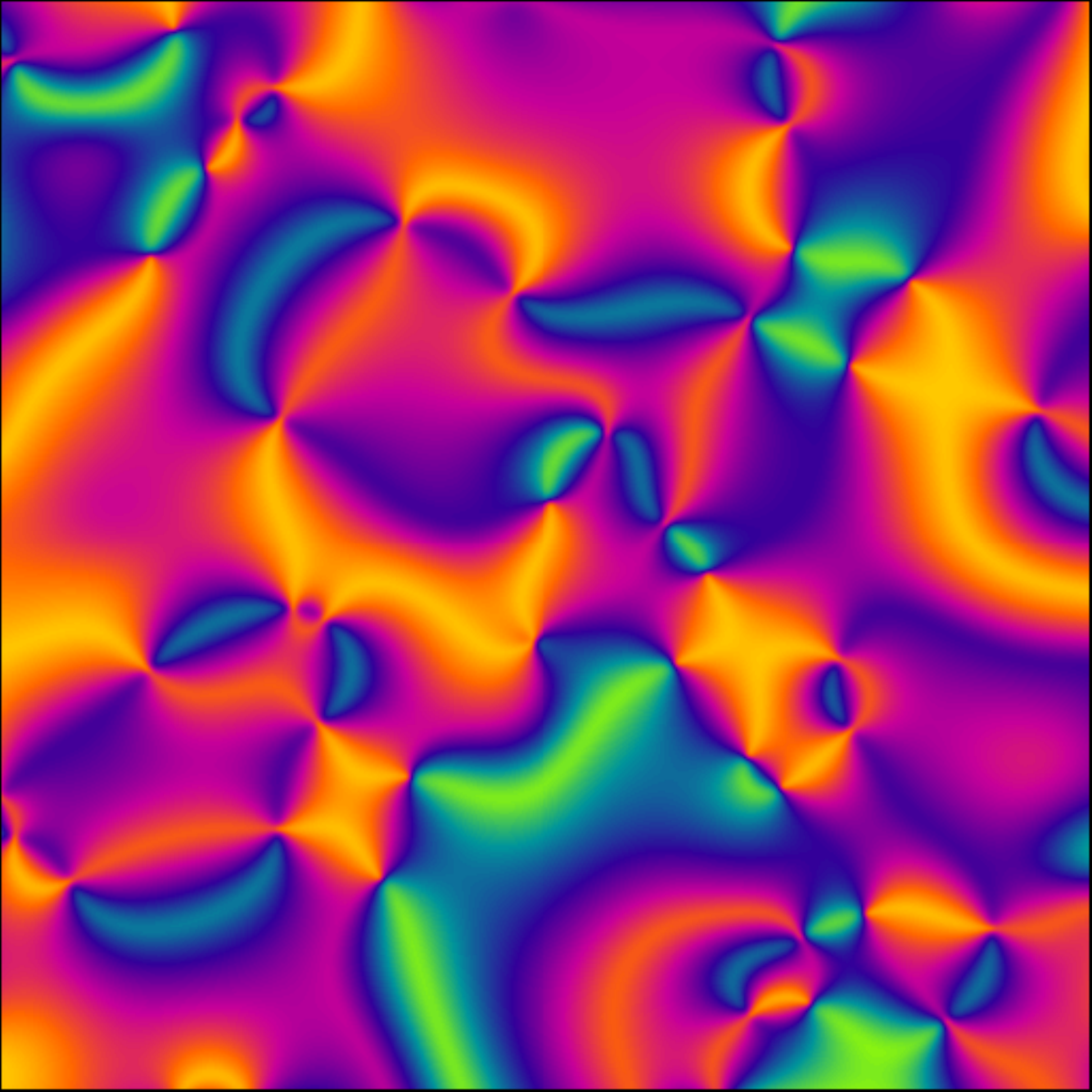}
\hspace{2cm}
\includegraphics[width= 4.2cm]{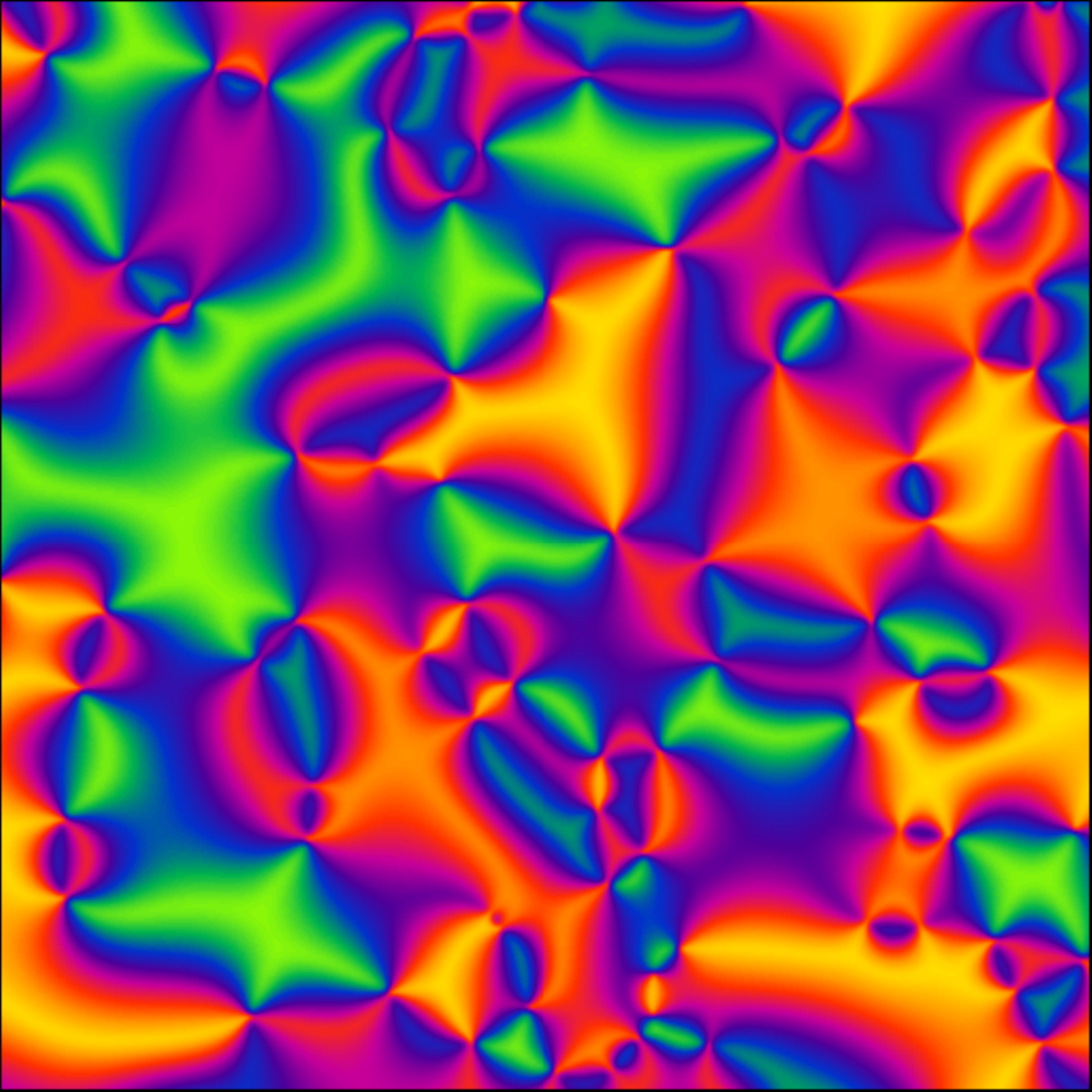}
\caption{Snapshots obtained from two-dimensional
$1024^2$ mean field simulations of the $N=6$ (left
panels) and $N=8$ (right panels) models.}
\label{fig6}
\end{figure}

Figure~\ref{fig6} depicts two snapshots taken from $1024^2$ mean field
network simulations with periodic boundary conditions of the $N=6$
(left panel) and $N=8$ (right panel) models. Initial conditions with
$\phi_i=1$ if $i=s$ and $\phi_i=0$ if $i\neq s$ were set at each grid
point ($\phi_0$ was set to zero at every grid point). Here $s$ is a
species drawn randomly at each grid point.

\begin{figure}[t]
\centering
\includegraphics[width= 4.2cm]{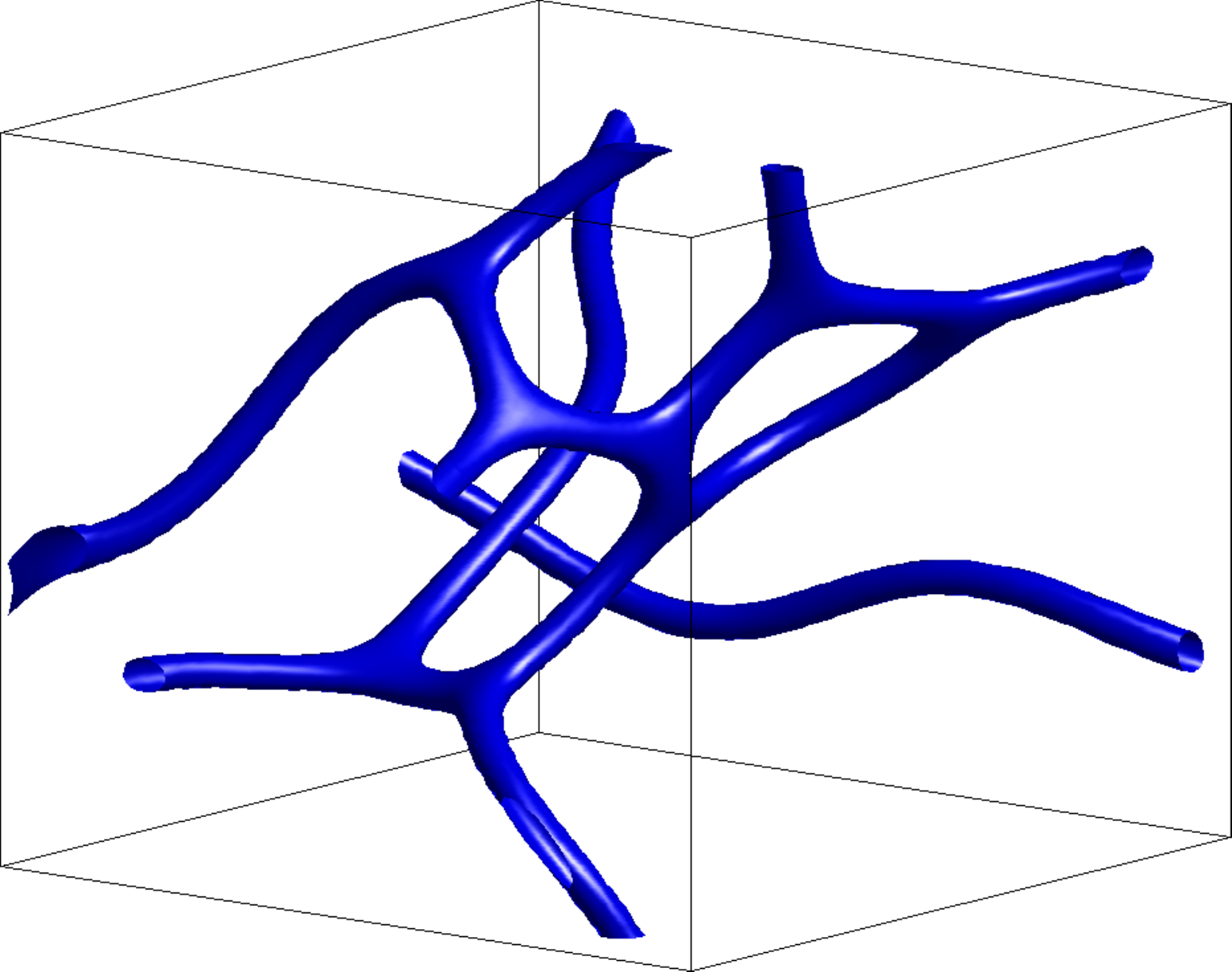}
\hspace{2cm}
\includegraphics[width= 4.2cm]{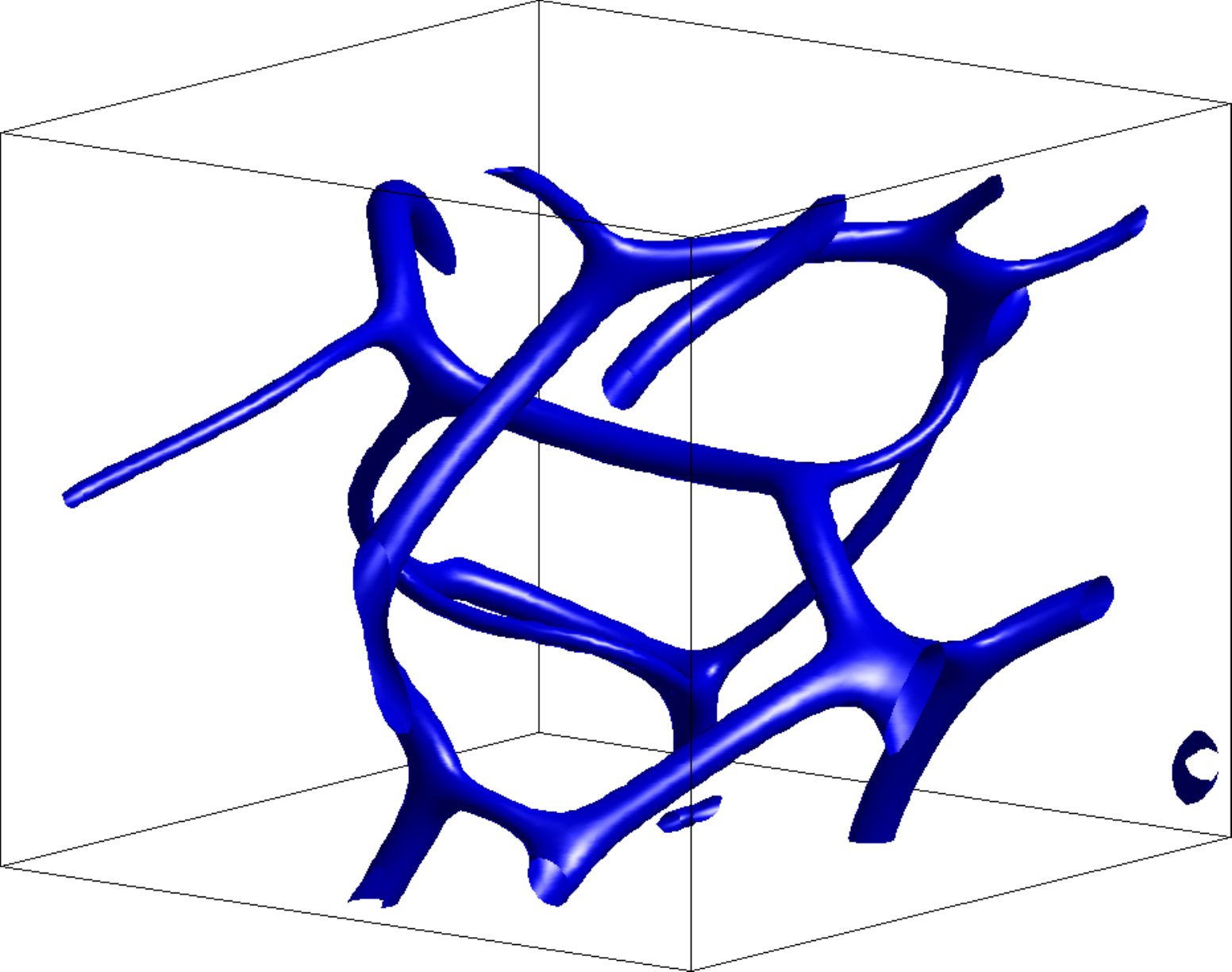}
\caption{Snapshots taken from three-dimensional 
mean field simulations of the $N=6$ (left panel) and $N=8$ (right
panel) models. The figures represent $64^3$ regions of $256^3$
simulations.}
\label{fig7}
\end{figure}

The results provided by the mean field simulations (Fig.~\ref{fig6})
are similar to those obtained from the stochastic network evolution
(Fig.~\ref{fig3}), except for the noise (the color scheme is the same
in both figures). This correspondence also occurs in the
three-dimensional simulations. In particular, the macroscopic dynamics
depicted in the snapshot shown in Fig.~\ref{fig7} obtained from
three-dimensional $256^3$ mean field theory simulations of the $N=6$
and $N=8$ models, is similar (again, except for the noise) to that
shown in Fig.~\ref{fig4}.

\subsection{Defect profile}

Let us now consider the defect profiles, i.e., the stationary number
density of empty spaces in and around the defect cores. In other
words, let us study the spatial distribution of $\phi_0$ around a
defect center, considering that the defect has radial symmetry and is
centered at $r=0$.

Furthermore, we aim to understand the dependence of the defect profile
on the parameters ($m$, $p$, $r$). To this purpose, we consider three
different models characterized by a domination of motion [model $M$:
($m$, $p$, $r$)=(0.8, 0.1, 0.1)], predation [model $P$: ($m$, $p$,
$r$)= (0.1, 0.8, 0.1)] or reproduction [model $R$: ($m$, $p$, $r$) =
(0.1, 0.1, 0.8)] interactions.

\begin{figure}[ht]
\centering
\includegraphics[scale=0.9]{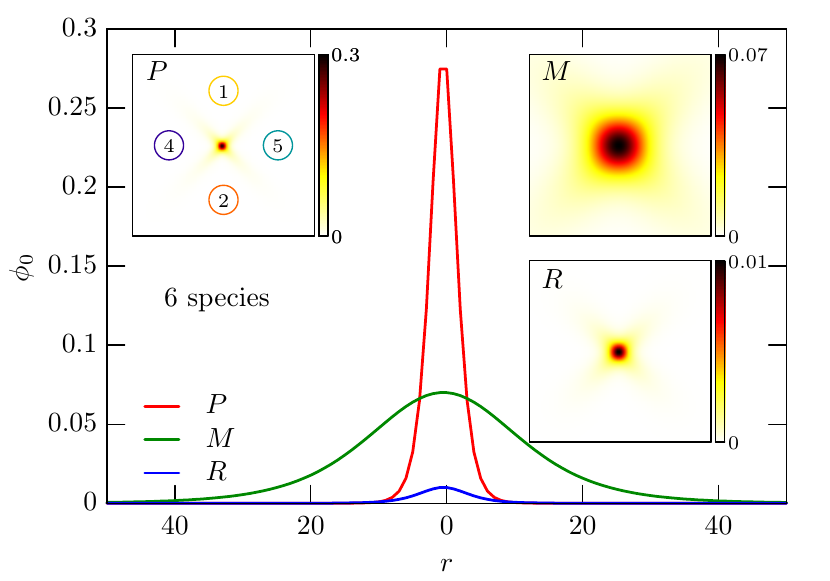}
\includegraphics[scale=0.9]{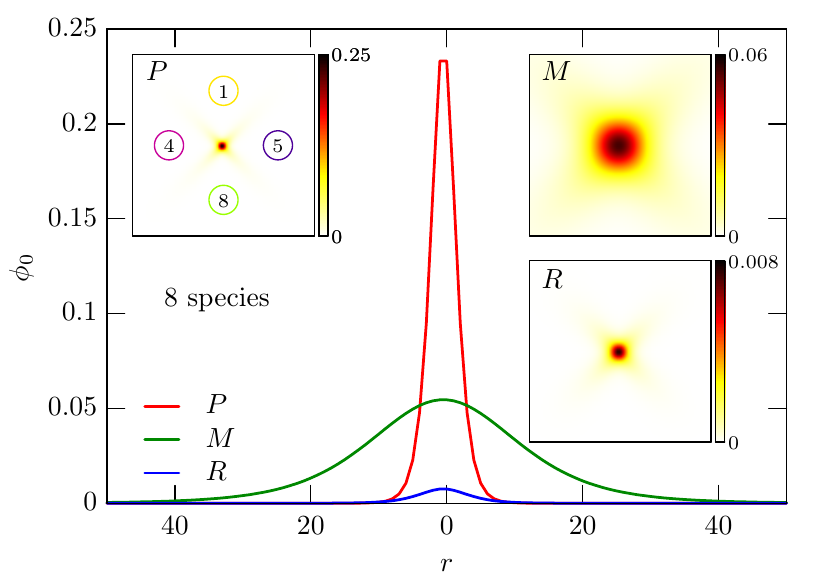}
\includegraphics[scale=0.9]{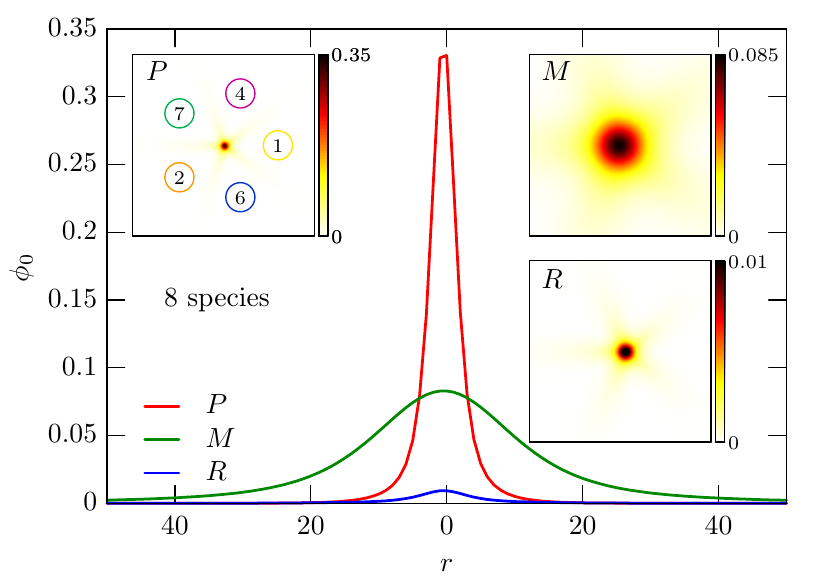}
\caption{Defect profiles obtained for defects with
four and five species from mean field simulations for $N=6$ (left
panel) and $N=8$ (right and lower panels). The inset plots represent
the disposition of the species around the defect cores.}
\label{fig8}
\end{figure}

Figure ~\ref{fig8} shows the value of $\phi_0$, as a function of the
distance $r$ to the defect core, obtained for defects associated with
four and five species from mean field simulations of the $N=6$ (upper
panel) and $N=8$ (middle and lower panels) models. The disposition of
the species around the defect cores is shown in the inset plots.
The numerical results show a strong dependence of the defect profile
on the model parameters. Specifically, the defect profile height is
larger in the case of model P, dominated by predation interactions. On
the contrary, a larger reproduction rate leads to a smaller number
density of empty spaces (model $R$). Finally, a larger mobility
parameter leads to the broadening of the defect profile, since the
individuals are more likely to move outside the core of the defect
into enemy territory (model $M$).

Figure ~\ref{fig8} also shows that, in the model with $N=8$ species
and for fixed ($m$, $p$, $r$), the defect is broader when the number
of species composing the defect is increased from $4$ to $5$.

\subsection{String loop}
\label{loop}

We now investigate the collapse of a circular string loop using mean
field theory simulations in a cubic lattice. To identify the string we
define a new variable $\varphi({\vec r},t)\equiv {\rm
max}(\phi_0({\vec r},t))-\phi_0^{c},0)$. Here, $\phi_0^{c}$ represents
a threshold which guarantees that only grid points with a high number
density of empty sites, close to the core of the string, are
identified as belonging to the string. The average number density of
empty sites associated to the string is then defined by
\begin{equation}
	\rho(t)=\frac1{{\cal N}^2}\sum_{\vec r}\varphi({\vec r}, t)\,.
	\label{rhodef}
\end{equation}
We recall that the average number of empty sites per unit string
length ($\mu$) does not change significantly with time. Therefore, the
loop perimeter is proportional to $\rho(t)$ and the area of the string
loop $a(t)$ evolves proportionally to $\rho^2(t)$.

\begin{figure}[t]
\centering
\includegraphics[scale=0.97]{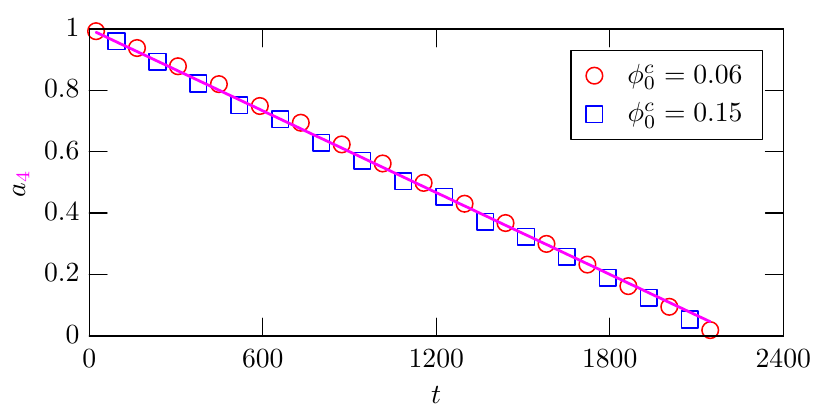}
\includegraphics[scale=0.97]{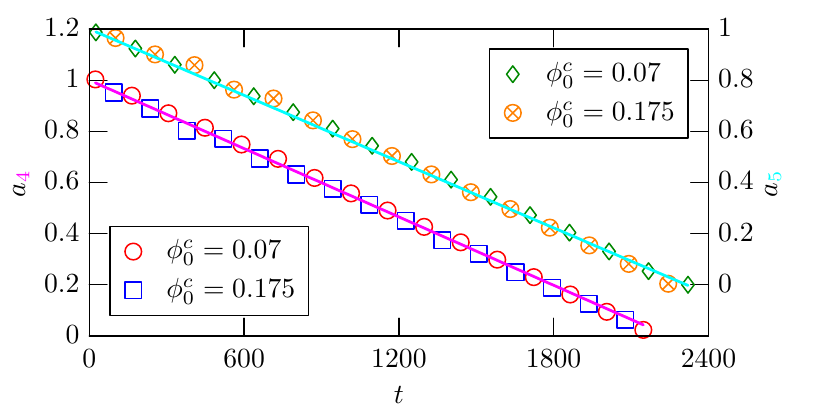}
\caption{Mean field evolution of the areas of string
loops in the $N=6$ (left panel) and $N=8$ (right panel) models,
computed by taking different threshold values. Note that $a_4(t)$ and
$a_5(t)$ represent the areas of type I and type II string loops,
respectively.}
\label{fig9}
\end{figure}

The time evolution of the area $a(t)$ of the circle enclosed by the
loop is determined using mean field theory simulations in a $256^3$
cubic lattice. We run simulations for two different threshold values
in order to ensure the accuracy and reliability of the results. The
upper panel of Fig. ~\ref{fig9} displays the evolution of the area of
a string loop formed by four distinct species $a_4(t)$ in the $N=6$
model. Note that the results are almost identical for both choices of
threshold values, $\phi_0^c = 0.06$ and $\phi_0^c = 0.15$, which
correspond approximately to $25\%$ and $60\%$ of the maximum of
$\phi_0$ ($\phi_0^{max}$) at the core of the string. Similar results
were found for the evolution of area of the circle enclosed by string
loops associated to four and five distinct species $a_4(t)$ and
$a_5(t)$ in the $N=8$ model (see lower panel of Fig. ~\ref{fig9}).
Fig. ~\ref{fig9} shows that the loop area decreases linearly in time
according to
\begin{equation}
a(t)=a_0\left(1-\frac{t}{t_c}\right)\ , 
	\label{a}
\end{equation}

where $t_c$ is the collapse time or, equivalently, that radius of
curvature decreases proportionally to $t^{1/2}$. A similar result has
been obtained in Ref.~\cite{PhysRevE.86.036112} for a different model
allowing for string networks without junctions.
\begin{figure}[t]
\centering
\includegraphics[scale=1.0]{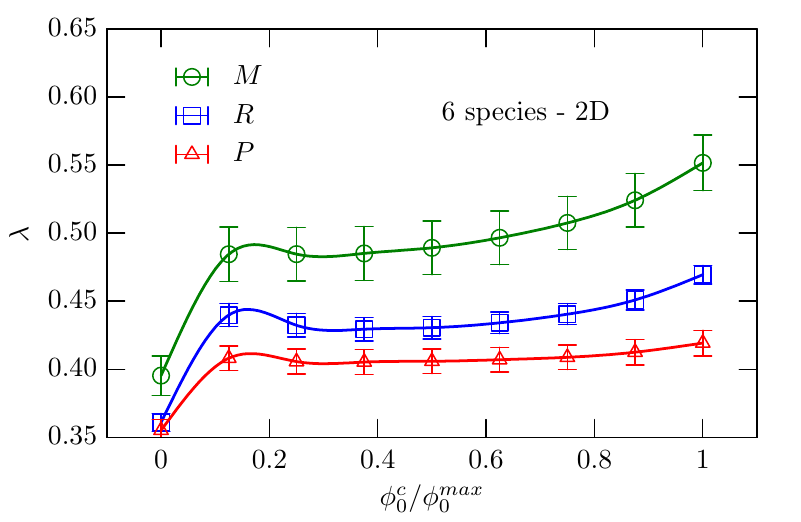}
\includegraphics[scale=1.0]{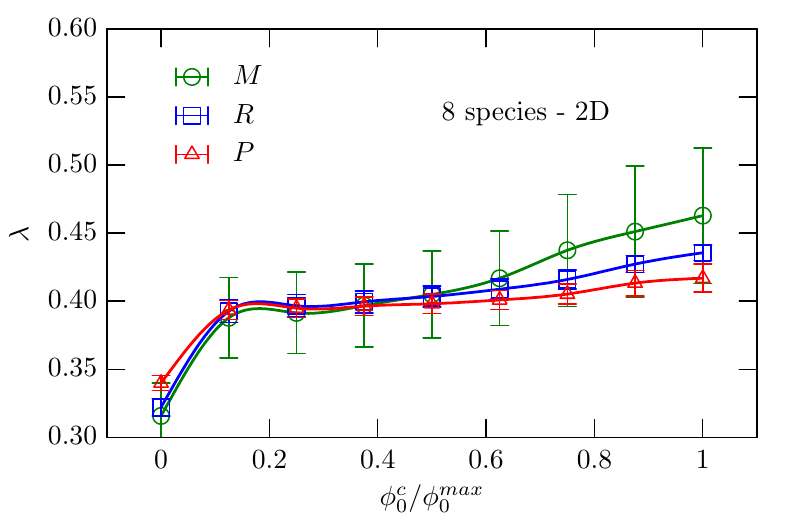}
\caption{The dependence of the scaling exponent
$\lambda$ on the threshold $\phi_0^{c}$ in a mean field simulation
with $N=6$ (left panel) and $N=8$ (right panel) species. The results
were obtained by carrying out $10$ simulations of $2048^2$
two-dimensional networks for a wide range of $\phi_0^{c}$, for models
$M$, $P$ and $R$. The error bars represent the standard deviation in
an ensemble of 10 simulations.}
\label{fig10}
\end{figure}
\begin{figure}[t]
\centering
\includegraphics[scale=0.97]{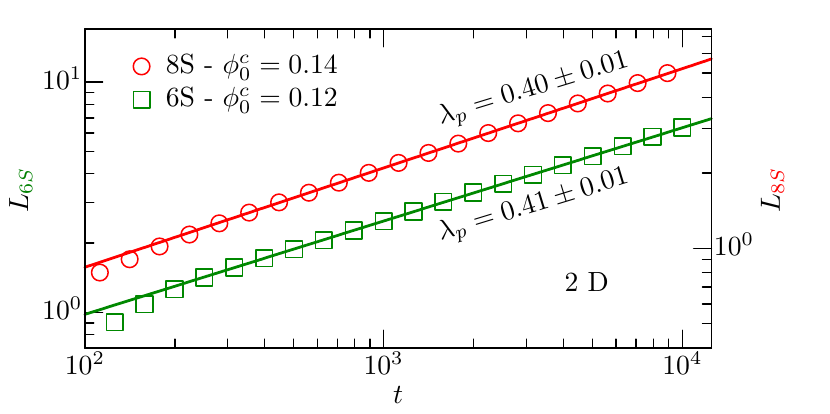}
\includegraphics[scale=0.97]{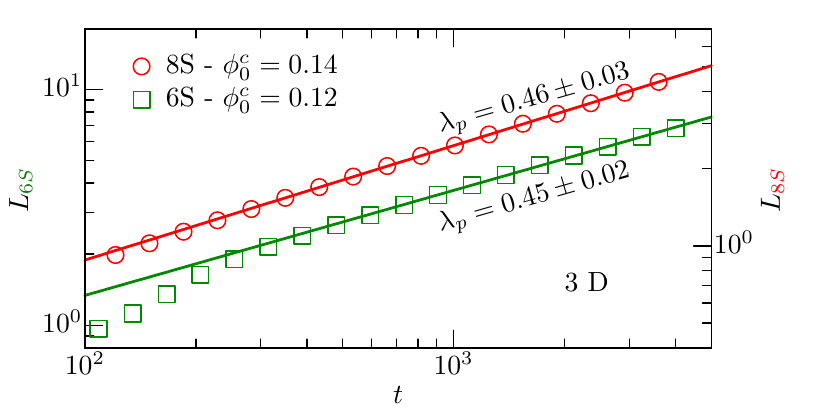}
\caption{Scaling behavior for the models with $N=6$ and
$N=8$ different species obtained using $2048^2$ two-dimensional (left
panel) and $256^3$ three-dimensional (right panel) mean field
numerical simulations.}
\label{fig11}
\end{figure}

\section{Scaling behavior}
\label{sec:sc}

Finally, we consider the scaling behavior of the models investigated
in the previous sections using mean field theory simulations. To this
purpose, let us define the characteristic length $L$ of the defect
network as
\begin{equation}
	L = \sqrt{\frac{\mu}{\rho}} \propto \sqrt{\frac1{\rho}}\ ,
	\label{L}
\end{equation}
where $\mu$ is the average number of empty spaces associated with the
defects (per unit string length, in three spatial dimensions) and
$\rho$ is the average number density of empty sites associated to the
defect network defined by Eq. (\ref{rhodef}).

By fitting a scaling law $L \propto t^{\lambda}$, we constrain the
evolution of the characteristic length of the string network with time
for different sets of $10$ mean field simulations of the models $M$,
$R$ and $P$. Each simulation starts with different random initial
conditions.

The upper and lower panels of Fig. ~\ref{fig10} show the dependence of
$\lambda$ on the threshold $\phi_0^{c}$ for models $M$, $P$ and $R$,
for $N=6$ and $N=8$, respectively (in units of
$\phi_0^c/\phi_0^{max}$). For $N=8$, $\phi_0^{max}$ is taken as the
maximum of $\phi_0$ at the core of the string of type I. The error
bars represent the standard deviation in an ensemble of 10
simulations. Figure ~\ref{fig10} shows that if the $\phi_0^{c}$ is
between $20\%$ and $60\%$ of $\phi_0^{max}$, the scaling constant
$\lambda$ does not show a significant dependence on the threshold.
Outside this interval, this is no longer the case. For lower values of
$\phi_0^{c}$, this happens because low density regions far away from
the core are being taken as belonging to the defect. On the other
hand, the number of lattice points associated with the defect may
become too small for higher values of $\phi_0^{c}$.

The average evolution of $L$ with time $t$ was obtained for sets of 10
distinct two- and three-dimensional mean field network simulations
($2048^2$ and $256^3$) with random initial conditions. Fig.
~\ref{fig11} shows that the characteristic lengths $L_{6S}$ (6
species) and $L_{8S}$ (8 species) evolve in reasonable agreement with
the scaling law $L\propto t^\lambda$, with $\lambda = 1/2$,
characteristic of networks in which the dynamics is curvature driven.
Here, the points denote the average value of $L$ computed from the
simulation and the error bars provide information on the
root-mean-square deviation for each set of $10$ simulations. In all
cases the value of $L$ was normalized to unity at $t = 100$, the
parameters ($m$, $p$, $r$) were set to (0.10, 0.80, 0.10) and the
threshold $\phi_0^{c}$ was fixed at $40\%$ of $\phi_0^{max}$ (for
$N=8$ the treshold $\phi_0^{c}$ was obtained by considering the value
of $\phi_0^{max}$ obtained for type II strings).

These results are consistent with those obtained in
Ref.~\cite{Avelino2014393} for models allowing for string networks
without junctions in three spatial dimensions. A similar behavior may
also be found in other physical systems, in particular in the case of
curvature driven dynamics of string networks in condensed matter.

\section{Comments and Conclusions}
\label{sec:end}

In this work we have shown that there are specific sub-classes, with
an even number of species, of a more general family of May-Leonard
models which lead to the formation of cosmic string networks with
junctions, associated to regions with a high concentration of empty
spaces. We have investigated the dynamics of these networks using
stochastic and mean field network simulations, assuming that the
predation, reproduction and mobility probabilities are constant in
space and time. We have found that the presence of junctions does not
have a significant impact on the scaling behaviour of the
characteristic macroscopic scale of the network $L$ with the physical
time $t$, showing that it grows roughly proportional to $t^{1/2}$. We
have also shown that our results are not strongly dependent on the
specific values of the mobility, predation or reproduction
probabilities.

Our findings are expected to be relevant not only for the evolution of
biological populations in three spatial dimensions but also for the
understanding of the evolution of cosmic string networks with
junctions in other contexts. In particular, in
\cite{PhysRevD.79.085007} one investigates bifurcation and pattern
changing in the relativistic regime, suggesting how to solve the
cosmological domain wall problem. Also, in \cite{Avelino2010} it has
been shown that, although the evolution of the characteristic
macroscopic length and velocity of interface networks with physical
time in relativistic and non-relativistic regimes is very different,
single snapshots with the same characteristic scale do not clearly
differentiate between these two regimes. Hence, our results may also
be relevant for the understanding of the dynamical behaviour of
complex relativistic defect networks with junctions in a cosmological
setting (e.g. cosmic superstrings \cite{2004JHEP...06..013C,
2005JHEP...10..013J, PhysRevD.94.063529}).

\section*{Acknowledgements}
\label{thanks}

We thank CAPES, CNPq, CNPq/Fapern, and FCT-Portugal for financial
support. The work of PPA was supported by Funda\c c\~ao para a
Ci\^encia e a Tecnologia (FCT) through the Investigador FCT contract
of reference IF/00863/2012 and POPH/FSE (EC) by FEDER funding through
the program {\it Programa Operacional de Factores de Competitividade},
COMPETE.

\bibliographystyle{elsarticle-num}
\bibliography{ablmo}

\end{document}